\documentclass[aps,amsmath,amssymb,prl,twocolumn,showpacs,nofootinbib,flushbottom]{revtex4-1}
\usepackage{epsfig}

\usepackage[usenames]{color}\newcommand{\bhx}{\mathbf{\hat x}}
\newcommand{\bhy}{\hat y}
\newcommand{\bhz}{ \hat z}

\newcommand{\be}{\mathbf{e}}

\newcommand{\bM}{\mathbf{m}}
\newcommand{\bP}{\mathbf{P}}

\newcommand{\bhn}{\bf{\hat n}}

\newcommand{\bnabla}{\mathbf\nabla}
\newcommand{\br}{\mathbf r}

\newcommand{\bq}{\mathbf q}

\newcommand{\bn}{\mathbf n}

\begin{document}

\title{Vortex domain walls in  helical magnets}
\author{Fuxiang Li$^1$}\author{T. Nattermann$^2$}\author{V.L. Pokrovsky $^{1,3}$}\affiliation{$^1$Department of Physics, Texas A\&M University, College Station, Texas77843-4242}\affiliation{$^2$Institut f\"ur Theoretische Physik, Universit\"at zu K\"oln, D-50937 K\"oln, Germany}
\affiliation{$^3$Landau Institute for Theoretical Physics, Chernogolovka, Moscow District,
142432, Russia}
\date{\today}

\begin{abstract}
We show that helical magnets exhibit a  non-trivial type of domain wall consisting  of  a regular array of vortex lines, except of a few distinguished
orientations.
This result follows from topological consideration and is independent of the microscopic models. We used simple models to calculate
the shape and energetics of vortex walls in centrosymmetric and non-centrosymmetric crystals. 
Vortices are strongly anisotropic, deviating  from the conventional Berezinskii-Kosterlitz-Thouless form. The width of the domain walls depend  only weakly  on the magnetic  anisotropy, in contrast to ferromagnets and antiferromagnets. 
We show that vortex walls can be driven by external currents and in multi-ferroics also  by electric fields. \end{abstract}


\maketitle

{\it Introduction.}--- The structure of domain walls  (DWs) determines to a large extent the properties of  magnetic materials, in particular their hardness and switching behavior,  it represents an essential ingredient of spintronics   \cite{Hubert74,Parkin+08}. 
Common   DWs are of Bloch and Ne\'el types  in which the magnetization rotates  around a fixed axis,  
giving rise to a { one}-dimensional magnetization profile \cite{Bloch32,
Neel48}. 
Two-dimensional vortex wall configurations can appear in restricted geometries as a result of the competition of  stray field, exchange  and anisotropy energy \cite{Hubert74}.  
{ The } more difficult problem of DWs in helical magnets has not yet been solved.

Here we show that  DWs in helical magnets are fundamentally different from Bloch and Ne\'el walls. 
They are  generically characterized by a two-dimensional pattern. For almost all orientations of the DW they contain a regular lattice of vortex singularities. However DWs of few exceptional orientations, determined by symmetry, are free of vortices  and maximally stable.
Though  DWs do not exist without anisotropy, their width and energy depend only weakly on the anisotropy strength. 
 Similar to other topological defects \cite{Thiele73,Berger84,Zhang+04,Tatara+08}, vortex DWs can be driven by electric currents.
In multi-ferroics  vortices are electrically charged, allowing manipulation of magnetic DWs by electric fields \cite{Mostovoy06,Cheong+07,Arima11}. 

\begin{figure}[h]   
 \centering\includegraphics[width=7cm]{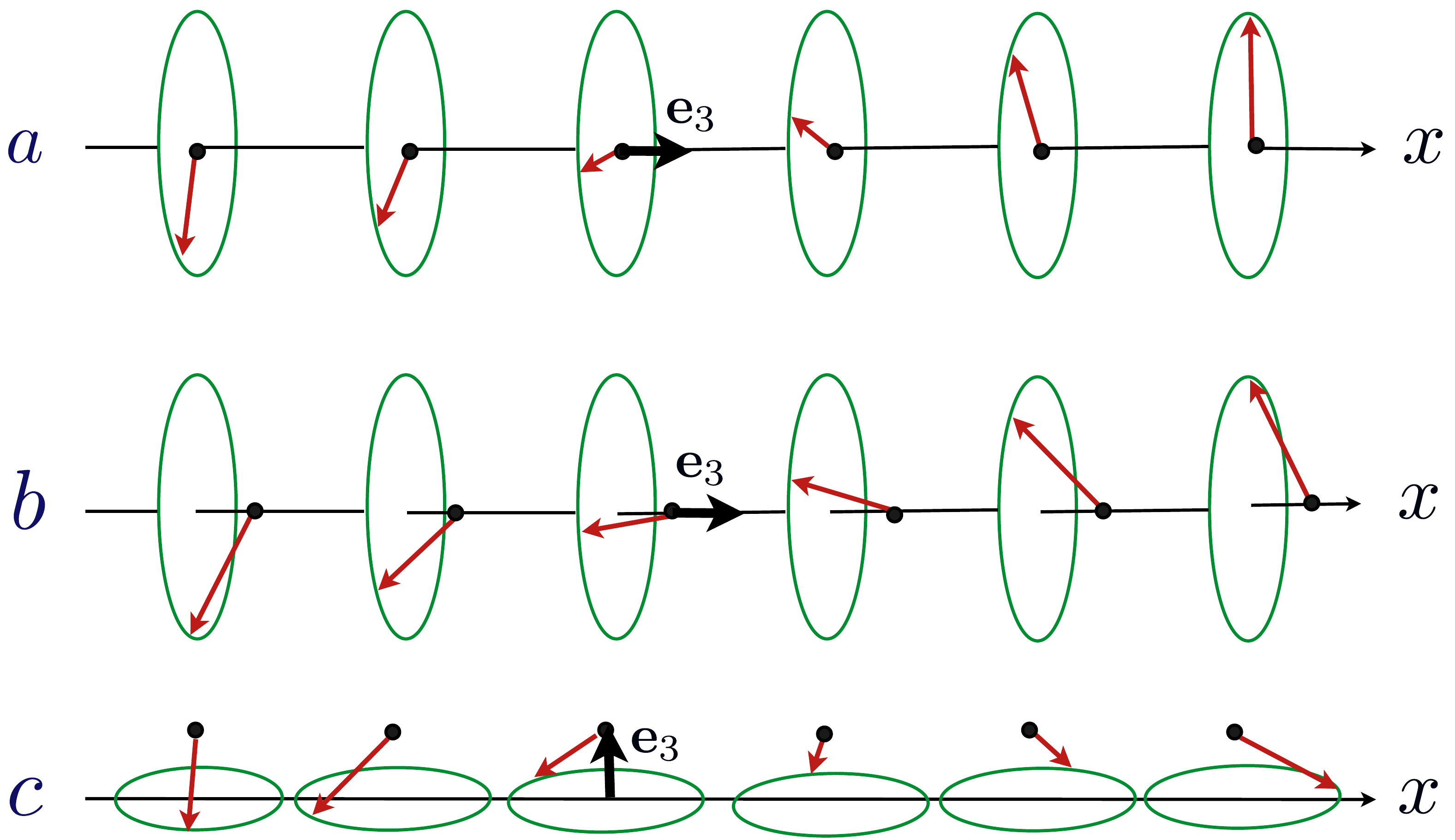} 
 \caption{ Different types of helical ordering. 
 (a) The magnetization rotates in a plane perpendicular to the helical (x-) axis as in Tb, Dy, Ho. 
 (b) Conical phase with non-zero $m_3$-component of the magnetization as in Ho below 19K. 
 (c) The magnetization rotates in a plane   parallel to the helical axis as inTbMnO$_3$.} 
 \label{order} 
 \end{figure}

Helical magnets  exhibit a screw-like periodic spin pattern intermediate between ferromagnets and antiferromagnets. 
Examples of such structures are shown in Fig.\ref{order}. 
In addition to  time reversal symmetry, in helical magnets  the space inversion symmetry  is  broken \cite{Kleman70}, either spontaneously in centrosymmetric crystals,  or enforced by the symmetry of the crystalline lattice in non-centrosymmetric crystals. 
The magnetization $\bM$ in these structures rotates around a fixed axis when the coordinate along a fixed direction, generally not coinciding with the rotation axis, changes. Further we denote the projection of the magnetization to the rotation axis $m_3$, its rotating projection to the perpendicular plane as $\bM_{\perp}
$ and assume that $\bM^2 =1$. 
The angle of rotation is $\phi$.
 
{\it Centrosymmetric case.---} 
We begin with the { centrosymmetric case}, since it is simpler and includes already
many features discussed in this article.
Prominent experimental realizations   are frustrated antiferromagnets in rare earth metals 
 Tb, Dy, Ho \cite{Holmium_Koehler+66,Jensen+91},  their alloys and compounds { RMnO$_3$  R $\in $ \{Y, Tb, Dy\}  { \cite{Kimura+08}}, R$_2$Mn$_2$O$_5$, R$\in$ \{Tb, Bi\}, as well as  Ni$_3$V$_2$O$_8$ and LiCu$_2$O$_2$  } \cite{Kimura+08,Chapon+04}.
The helical magnetic order originates in these materials from the indirect RKKY exchange  
 which  results in a  competing  nearest neighbor ferromagnetic ($J>0$) and next nearest neighbor antiferromagnetic ($J'<0$) interaction along the helical axis 
  \cite{Jensen+91,deGennes62,Harris+06}. %
  The corresponding Ginzburg-Landau Hamiltonian  then reads \cite{Hubert74}  
\begin{align}
 \label{eq:Hamiltonian1}
{\cal H}_{c}=\frac{J}{2a}\int_{\br}\Big\{-\frac{\theta^2}{2}
 (\partial_x \bM_{\perp} )^2&+ \frac{a^2}{4} (\partial_x^2 \bM_{\perp} )^2+\\+(\bnabla_{\perp} \bM)^2+\nonumber(\partial_x m_3)^2&+\gamma^2(m_3^{2}+\tau\cos^{2}\vartheta_0)^2\Big\},
\end{align}
where $\int_{\br}=\int {d^3r}$, $\bnabla_{\perp}={\bhy}\partial_y+\bhz\partial_z$, 
and $a$ is   the lattice constant. 
$\theta=\arccos(J/4|J'|)$ denotes the angle between spins in neighboring layers. 
 The continuum approach is valid for $\theta\ll 1$.  $\theta$ can be diminished  to zero under uniaxial pressure \cite{Andrianov+00}.  
The last term in (\ref{eq:Hamiltonian1}) is an interpolation that fixes the spins either in-plane, $m_3=0$  at $\tau=(T-T_0)/T_0>0$, as  in Tb, Dy, Ho and TbMnO$_3$, or on a cone with angle $\vartheta_0$ for $\tau<0$, as  in Ho below $T_0=19K$ \cite{Lang+04} ($\vartheta_0\approx 1.56$ \cite{Jensen+91}). 
$\gamma a\approx 0.625$ for Ho and $\gamma a=0.17$ for Tb\cite{Jensen+91}.
 The ground state of  (1) has  a helical structure with $\phi=qx$: 
 \begin{equation}
 \bM=|m_{\perp}|\left( \be_1\cos qx+\chi\be_2\sin qx\right)+\zeta m_3\be_3
  \end{equation} 
  where $q=\theta/a$  (see Fig.\ref{order}a). $\chi=\pm1$ and $\zeta=\pm1$ describe the chirality and conicity of the solution, respectively. 
  The rotation axis  $\be_3$ may be parallel to the helical axis $\bhx$, as in Tb, Dy, Ho, or perpendicular to it,   as in  TbMnO$_3$ (see Fig.\ref{order}c). 
Because of its space inversion symmetry, (1)  is a   generic model  for any centrosymmetric helical magnet. 
In centrosymmetric helical magnets where the star of modulation vectors includes 3 vectors, like in  CuCrO$_2$ \cite{Arima11,Frontzek+11}, a slightly  more complicated model has to be used, but the main conclusions of our analysis remain valid also in this case.

{\it Domain walls and vortices.---}DWs separate half spaces with different values of $\zeta$ or $\chi$ or both.
We consider here only walls with different $\chi$ since domain walls between phases with  different $\zeta$, but the same value of $\chi$, are of  Ising type and well studied. 
 \begin{figure}[htbp]    
 \centering\includegraphics[width=8.7cm]{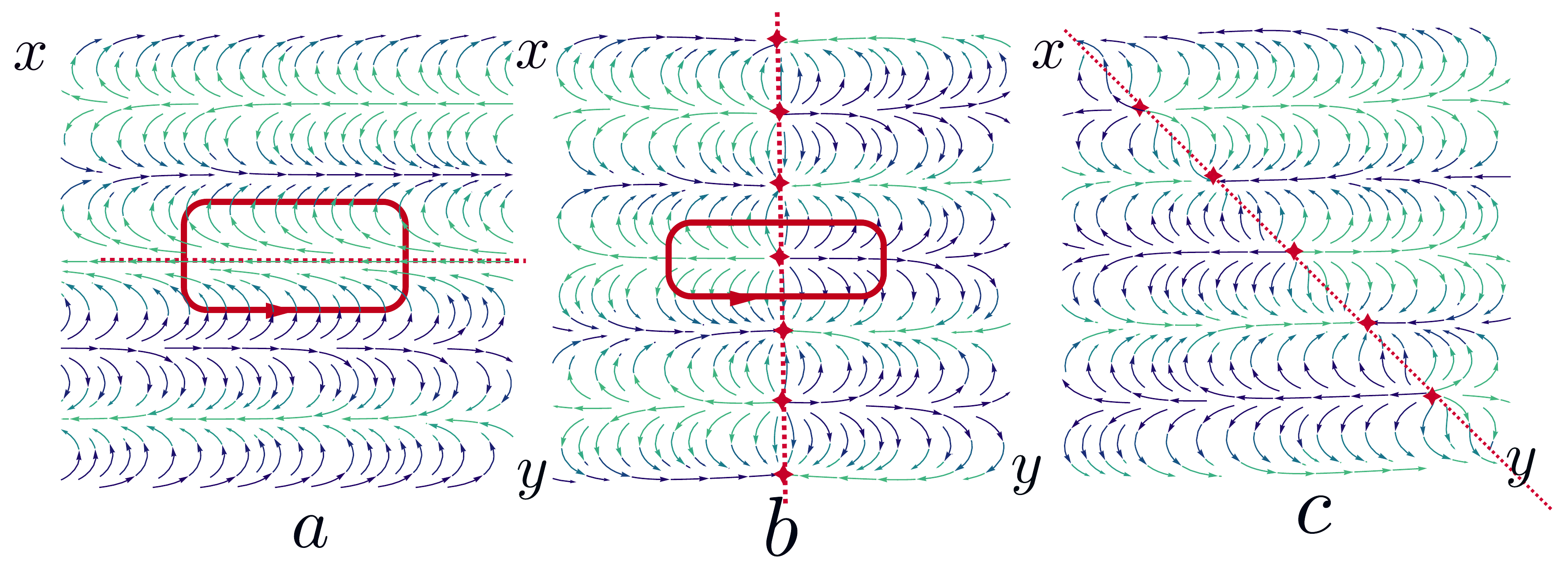}
 \caption{DWs  in centrosymmetric helical magnets. 
 Cross section parallel to the  $xy-$plane of  (a)  a Hubert wall,  (b) a vortex wall parallel to the helical axis in a system where   the magnetization rotates in the $x$-$y$ plane, (c) a vortex wall tilted with respect to the helical axis.  
 The arrows denote the orientation of $\bM$. 
 For systems where $\bM$ is confined to the  $yz$-plane,  $\bM$ have been rotated by $\pi/2$ for better visibility. 
 The red contour is described in the text } 
 \label{fig:3walls,planar}
  \end{figure}
A  wall whose   normal  $\bhn$ is parallel to the helical axis, $\bhn\cdot\bhx=1$,  has been  studied by Hubert  \cite{Hubert74,Melnichuk+02}.
In such a wall the derivative of the rotation phase $\partial_x \phi$ changes smoothly   from $-q$ to $q$ over a distance $\sim 1/q$ (see Fig.\ref{fig:3walls,planar}a).
Its surface tension  $\sigma_H\sim (J/a^2)|\theta|^3 $ is small for small $\theta$.
Walls of different orientation were not yet studied  theoretically, although seen in experiment, e.g. in Ho by circular polarized x-rays \cite{Lang+04}. 
We consider first  a wall in the $xz$-plane whose normal $\bhn$ is perpendicular to $\bhx$.  
Since both domains have the same pitch, the magnetization is periodic along $x$-axis with the period $2\pi/q$. 
Circulating counterclockwise along  a closed  contour ${\cal C}$ in the $xy$-plane formed by two horizontal lines at $x=N\pi/q$ and $x=(N+N_v)\pi/q$ with $N$ and $N_v$ being integers and two vertical lines connecting the horizontal ones far from the wall (see the red contour in Fig.\ref{fig:3walls,planar}b),  
an observer sees the  change of phase  $2\pi N_v$.
A  similar contour $\cal C$ enclosing a Hubert wall  gives $N_v=0$. 
We note that this argument { is} purely topological and  not limited to the particular Hamiltonian (\ref{eq:Hamiltonian1}). 
In the case of six modulation vectors $\pm \bq_i, i=1,2,3$, as in  CuCrO$_2$, in addition to the $\pm\bq_i$ DWs considered here, also DWs between $\bq_i,\bq_j$ phases ($i\neq j$)  appear, similar to those discussed below  for the non-centrosymmetric case.

 { Vortices   are 
 saddle point configurations} of the Hamiltonian (\ref{eq:Hamiltonian1}). For  $\gamma a\gg 1$ they obey the equation
\begin{eqnarray}
\label{eq:saddlepoint}
\left\{4\bnabla_{\perp}^2+{a^2}\left[6(\partial_x\phi)^2-2q^2-\partial_x^2\right]\partial_x^2\right\}\phi=0.
\end{eqnarray}
Vortex lines parallel to $\bhx$ have the standard Kosterlitz-Thouless form  \cite{Kosterlitz+73}.
The same applies to vortex lines perpendicular to $\bhx$ on scales much larger than $q^{-1} $ where $(\partial_x\phi)^2\approx q^2$ and hence  eq. (\ref{eq:saddlepoint}) becomes   Laplace's equation.  
On smaller scales, instead of solving (\ref{eq:saddlepoint}) exactly, we use a variational Ansatz
$\phi({\bf r})=\arctan( \lambda z/{x})$,
 where $\lambda$ is a variational parameter to be found from the energy minimization. 
 It gives  
$\lambda^2(r) = \theta^2+5/(64\ln (r/a))$ where $r^2=x^2+\lambda^2z^2$.
The vortex energy per unit length is
\begin{eqnarray}\label{eq:vortexenergy}
\varepsilon_v(r)=\frac{\pi J}{a}\ln ^{1/2} (r/a)\left[{5\over 64}+\theta^2\ln (r/a)\right]^{1/2}.
\end{eqnarray}
 (\ref{eq:vortexenergy}) describes the crossover from the conventional Kosterlitz-Thouless behavior $\sim\ln(r/a)$  at distances $r>r_c=a \exp[{5/(64\theta^2)}]$ to a $[\ln (r/a)]^{1/2}$ behavior at scales $r<r_c$.  

So far we assumed that $\gamma a\gg 1$ and hence the spins are confined at a fixed value of $m_3$.
However for $\gamma a<1$ in the vortex center, i.e. for
$r\lesssim r_{\gamma}=\gamma^{-1}(1+\tau\cos^2\vartheta_0)^{-1}\left|\ln  (\gamma a)\right|^{1/2},
$
spins align parallel to the $\be_3$-axis to save energy. Thus  $m_3\zeta=\pm1$, i.e. the vortex forms a meron \cite{Senthil+04}.
%
 %
 Vortices in the DW have the same vorticity $\pm1$ and are equidistant with the spacing $\pi/q$ forming a { vortex fence}. %
The energy per unit area of the vortex DW is 
$ \sigma_v=(\sqrt{5} J/4a^2)|\theta||\ln|\theta||^{1/2}\gg \sigma_H$.

 A DW of  general orientation with  $\bhn\cdot{\mathbf {\hat x}}=\cos\alpha$ consists of a periodic chain 
of vortices  perpendicular to the helical axis and the normal to the DW (Fig.\ref{fig:3walls,planar}c). 
 For $\alpha$ close to $0$ the wall can be treated as pieces of Hubert walls separated by vortex steps of the height $\pi/q$ and length $(\pi/q)/|\tan{\alpha}|$, 
 giving rise to a vortex staircase. The energy per unit area of such a wall is approximately equal to to $\varepsilon_v(q^{-1} q|\sin\alpha|/\pi$. At any $\alpha\neq 0$, it is larger than the energy of the Hubert wall.
 
{\it Non-centrosymmetrics case.---} In  these systems
{  invariants 
violating  the space  but not time inversion symmetry are permitted. Those terms   appear   in  first order  perturbation theory in the spin-orbit coupling constant  $ g$ \, \cite{Dzya58,Moriya60}.
Experimental  examples   of non-centrosymmetric compounds are   MnSi  \cite{Muhlbauer+04}, 
Fe$_{1-x}$Co$_x$Si  \cite{Uchida+06} and FeGe   \cite{Uchida+08}.
The magnetic anisotropy in crystals with cubic symmetry is of the order $g^4$.
The phenomenological Ginzburg-Landau  functional for the magnetization $\bM$ has been derived in detail in  \cite{Ho+10} and takes the form
\begin{equation}
\label{GL-DM}
{\cal H}_{n}=\frac{J}{a}\int_{\br}\left\{(\bnabla\bM)^2+2g\bM\,(\bnabla\times\bM)+
v\sum_{i=1}^3m_{i}^4\right\}.
\end{equation}
Here we ignored other terms representing the cubic anisotropy since they do not influence our results  qualitatively. 
For $v=0$ the  minimum of energy (\ref{GL-DM}) is given by a planar chiral structure,
${\bM}({\bf r}) = \be_1 \cos{\bf q\, r}+\be_2\sin{\bf q\, r}$,
where $\bf{q}$ is the wave vector of the helix and $\be_1$, $\be_2={\mathbf{\hat q}}\times \be_1$ and ${\mathbf{\hat q}}$ form a triad.
The direction of ${\bf q}$ is arbitrary, but its length $|{\mathbf q}|=g$ is fixed.
\begin{figure}[htbp]    
\centering
\includegraphics[width=8.5cm]{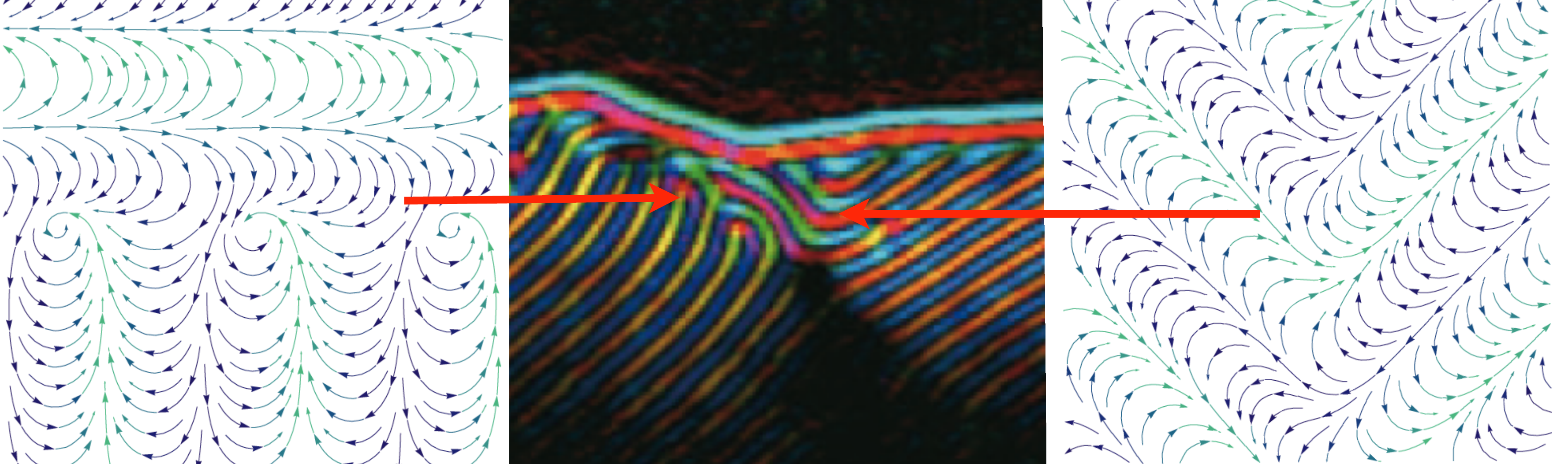} 
\caption{{ DWs in non-centrosymmetric helical magnets}.  Detail of Figure 1g of Ref.   \cite{Uchida+08} (center) showing two types of DWs in the ferromagnet FeGe, the left one includes  vortices, the right one is  vortex free. The panels are theoretically calculated DWs, right without vortices, left with vortices.}  
\label{fig:ncswalls}
\end{figure} 
Contrary to the centrosymmetric helices, states with  wave vectors $\bq$ and $-\bq$ describe the same magnetization reducing the degeneracy space  to $SO(3)/Z_2$  \cite{Volovik+}. 
Cubic anisotropy pins the helix direction  $\bf{q}$ either along one of the cube diagonals  or along one of the four-fold axis, depending on the sign of  $v$. 
DWs separate half spaces with different values of $\bq$. 
Since  $|v| \ll g^2$, one could expect,  in analogy with  ferromagnets, that the DW locally represents a helical structure whose wave vector slowly rotates pertaining its length constant. 
We will prove that such a configuration  does not exist. 
Indeed, 
the generalization of the equation  for the magnetization in a structure with slowly varying $\bq$ is
\begin{equation}\label{ansatz}
\bM(\br) = \be_1 \cos\phi(\br)+\be_2\sin\phi(\br),
\end{equation}
 where $\phi(\br)$ is an arbitrary function of coordinates.  
 $\be_1,\be_2,\bnabla\phi$ form a right triad. 
 The requirement of the constancy of the pitch implies
 $\left( \nabla\phi\right)^2 = q^2$,
which is the Hamilton-Jacobi equation for a free particle with the boundary conditions
$\nabla\phi\rightarrow{\bf q}_{1,2}$ at $x\rightarrow \mp\infty$.
Since a free particle conserves its momentum, the latter cannot be different in two different asymptotic regions.
Thus it is impossible to construct a DW between two different asymptotic values of the wave vector without changing its modulus between.
 The DW solution has a width determined by the only existing scale $1/q$ and the surface energy is independent of anisotropy $v$.

DWs whose plane is a bisector of the  asymptotic wave vectors $\bq_1$ and $\bq_2$ do not contain vortices. 
They are analogs of the Hubert DWs. Their surface tension has the order of magnitude $\sigma\sim Jg/a$.
DWs of any different orientation contain a chain of vortex lines for the same reason as in the centrosymmetric case (see Fig.\ref{fig:ncswalls},  right panel). 
 The vortex lines are located in the plane of the DW perpendicular to the  projection of either the vector $\bq_1-\bq_2\equiv2\bq_{-}$ or
$\bq_1+\bq_2\equiv 2\bq_{+}$ onto the domain plane depending on what configuration has lower energy. 
The vortex line spacings in the chain are equal to  $\ell_{\pm} =2\pi/\left|\bhn\times{\bf q}_{\pm}\right|$. 
Pictures of both vortex-free and vortex DWs based on a variational numerical calculations  are shown in Fig. \ref{fig:ncswalls}, together with the  experimental figure of FeGe   \cite{Uchida+08} displaying these structures. 
For numerical calculations we used (\ref{ansatz}) and the following ansatz (we write the answer
for the first choice of sign): %
\begin{equation}
\label{phase-var}
\phi(\br)={\br\bq_{+}}+{\bn\bq_{-}}w\ln\cosh\frac{\bn\br}{w}+\arctan\frac{\tan\psi_1}{\tanh\psi_2},
\end{equation}
where $\psi_1=\left[\br-\bn (\bn\br)\right]\bq_{-}$ and $\psi_2 = \left|\bn\times\bq_{-}\right|\bn\br$.
The last term in (\ref{phase-var}) is the contribution of the vortex array. 
It has the asymptotics $\pm\psi_1$. 
The second term does not have any singularity. 
It corresponds to the vortex-free DW when $\bn$ is parallel to $\bq_1-\bq_2$,
i.e. when the DW plane is the bisector of the vectors $\bq_1$ and
$\bq_2$. Its asymptotics are $\pm\left(\bn\br\right)\left(\bn\bq_{-}\right)$.
The asymptotic of the sum of the second and third terms is $\pm{\br}\bq_{-}$.
Together with
 the first term they tend asymptotically to $\bq_1\br$ above the domain wall and to $\bq_2\br$ below. The only variational parameter is $w$. %
 The surface tension of a vortex DW differs from that of the vortex-free DW by a factor $\sin\beta\ln(1/qa)$, where $\beta$ is the angle between $\bn$
 and $\bq_{\mp}$.
 Apart from a narrow interval of small $\beta$, this factor is larger than one. %
Because of their higher surface tension,  DWs carrying vortices  may be unstable with respect to formation of a zig-zag structure formed by vortex-free DWs.
 Zig-zag structures observed in experiments with  Fe$_{0.5}$Co$_{0.5}$Si \cite{Uchida+06} can be tentatively interpreted as arising from this instability. 
 The zig-zag structure is impossible in the helical magnets with uniaxial anisotropy since only one orientation of the vortex free DWs is allowed. 
This fact together with low stability of vortex-carrying DWs can serve as explanation of a disordered domain structure observed in Ho \cite{Lang+04}.

  {\it DW roughening.---}  
  Roughening  of  DWs occurs by formation of  { terraces} which condense at the roughening transition temperature \cite{Noziere}.
   For Hubert walls terraces   are encircled by  vortex rings of some length $L$. 
   Since their energy and entropy scale  as $\varepsilon_v(L)(L/a)$ and $ L/a$, respectively, Hubert walls  remain  asymptotically flat at increasing temperatures, slowing down their propagation.
    On the contrary, vortex walls are always rough, as seen also experimentally  \cite{Lang+04}. 

{\it Driven domain walls.---} 
 We assume that the spin  of a conduction electron follows adiabatically the magnetization  $\bM(\br)$. 
 This approximation is valid provided $|k_{\scriptsize F}^{\tiny\uparrow}-k_{\scriptsize F}^{\tiny\downarrow}|\gg q$.   %
 Here $k_{\scriptsize F}^{\tiny \uparrow\downarrow}$ is the Fermi momentum of the electrons with spin parallel or anti-parallel to $\bM$. 
 Thus,  electrons  experience a change of angular momentum. 
 Inversely,  the electron current $\bf j$ creates a reaction torque on $\bM$  driving  the magnetic texture with a force \cite{Thiele73,Berger84,Zhang+04,Tatara+08}
\begin{equation}\label{Berry1}
F_{\alpha}=\frac{\hbar}{2e}j_{\beta}\int_{\br}\left\{\bM\cdot\left(\partial_{\alpha}\bM\times\partial_{\beta}\bM\right)+\beta_{sf}\partial_{\beta}\bM\cdot\partial_{\alpha}\bM\right\}
\end{equation} 
The first term is the spin transfer torque \cite{Thiele73,Berger84} 
related to the Berry's curvature
$K_{\alpha}={\epsilon_{\alpha\beta\gamma}}\bM
\left(\partial_{\beta}\bM\times\partial_{\gamma}\bM\right).  
$ 
 For a single vortex its only non-zero component is parallel to the vortex lines and is given by  $
 2\pi m_3\zeta$.
  A weak field along the axis of rotation will order $\zeta$ of different merons.
 The force per unit area of the DW exerted by a current of density $j$ parallel to the wall due to the spin torque is  of the order $ m_3\zeta\theta({j}/{10^5Am^{-2}}) Nm^{-2}$.   
The second term results from the spin relaxation and is orthogonal to the first one. $\beta_{sf}$ is a dimensionless coefficient which depends on the specific relaxation mechanism \cite{Zhang+04,Tatara+08}. 
The pinning force density due to non-magnetic impurities of density $n_{i}$ can be estimated from the theory of collective pinning as $J\theta n_i /6\approx\theta({T_c}/{20K})({n_i}/{10^{17}cm^{-3}})Nm^{-2}$, which gives a critical current $j_c\approx 6\,10^7Am^{-2}$ for $n_i\approx 10^{19}cm^{-3}$.

{\it Multiferroics.---} In multiferroics the magnetization  can induce the electric polarization  \, \cite{Mostovoy06} 
 \begin{equation}
 \bP=\kappa\left[\bM(\bnabla\,\bM)-(\bM\,\bnabla)\bM\right],
  \end{equation} 
where $\kappa $ is some material constant. $\bf P$  is only non-zero if $\bM\,\bhx\neq0$ (as in TbMnO$_3$). The vortex structure in a helical DW induces a  ferroelectric DW, {in agreement with experiments \, \cite{Fiebig+02}}.  Hubert walls are uncharged whereas vortex lines carry an electric  charge  
$\rho=2\pi \kappa\left[\be_3\times\bhx\right]\bhn $ 
 per unit length.  This allows to move  magnetic DWs by an  external electric field.


To conclude, we have shown that DWs both in  centrosymmetric and non-centrosymmetric helical magnets consist of a regular array of vortex lines for almost all orientations except of a few that correspond to a minima of the surface energy. The helical DWs are generically 2-dimensional textures. They are charged in multi-ferroics and can be driven by electrical currents and fields.

\noindent{ The authors thank A. Abanov, T. Arima, K. Everschor, M. Kl\'eman, S. Korshunov, N. Nagaosa,  A. Rosch,  C. Sch\"ussler-Langeheine, G. E. Volovik, P.B. Wiegmann and M. Zirnbauer   for  useful discussions 
 and  
 Y.  Tokura for the  permission to reproduce his experimental figures.
This work has been supported by SFB 608 and by the DOE under
the grant DE-FG02-06ER 46278.}%

\end{document}